\documentclass[aps,prd,nofootinbib,superscriptaddress,preprintnumbers,reprint]{revtex4-1}
\pdfoutput=1 

\usepackage{graphicx}
\usepackage{amssymb}
\usepackage{amsmath}
\usepackage{bm}
\usepackage{hyperref}
\usepackage{datetime}
\usepackage{color}
\usepackage[utf8]{inputenc}
\usepackage[T1]{fontenc} % if needed
\usepackage{hepunits}

\renewcommand\Re{\operatorname{\mathfrak{Re}}}
\renewcommand\Im{\operatorname{\mathfrak{Im}}}

\newcommand{\Ht}{\widetilde{\mathcal{H}}}
\newcommand{\Et}{\widetilde{\mathcal{E}}}

\newcommand{\ImH}{\Im{\mathcal{H}}}
\newcommand{\ImE}{\Im{\mathcal{E}}}

\newcommand{\ImEt}{\Im{\widetilde{\mathcal{E}}}}

\newcommand{\ReH}{\Re{\mathcal{H}}}
\newcommand{\ReE}{\Re{\mathcal{E}}}
\newcommand{\ReHt}{\Re{\widetilde{\mathcal{H}}}}
\newcommand{\ReEt}{\Re{\widetilde{\mathcal{E}}}}

\newcommand{\xb}{x_{\mathrm{B}}}
\newcommand{\Q}{Q}

\begin{document}

\title{Separation of Quark Flavors using DVCS Data}

\author{Marija Čuić}
%\email{his@email}
\affiliation{Department of Physics, Faculty of Science, University of Zagreb, 10000 Zagreb, Croatia}

\author{Krešimir Kumerički}
%\email{his@email}
\affiliation{Department of Physics, Faculty of Science, University of Zagreb, 10000 Zagreb, Croatia}
\affiliation{Institüt für Theoretische Physik, Universität Regensburg, D-93040 Regensburg, Germany}

\author{Andreas Schäfer}
%\email{his@email}
\affiliation{Institüt für Theoretische Physik, Universität Regensburg, D-93040 Regensburg, Germany}

\begin{abstract}
Using the available data on deeply virtual Compton scattering (DVCS)
off protons and utilizing neural networks enhanced by
the dispersion relation constraint, we determine six out of eight leading Compton
form factors in the valence quark kinematic region. 
Furthermore, adding recent data on DVCS off neutrons, we separate contributions of
up and down quarks to the dominant form factor,
thus paving the way towards a three-dimensional picture of the nucleon.
\end{abstract}

%\date{\today, \currenttime}
\preprint{ZTF-EP-20-04}

\maketitle

%----------------------------------------------------------------------------------------
%	ARTICLE CONTENTS
%----------------------------------------------------------------------------------------

\section{Introduction}

Understanding the structure of hadrons in terms of their partonic
constituents (quarks and gluons) is one of the preeminent tasks
of modern hadron physics.
Over the years,
experiments like deep inelastic scattering (DIS) led to 
a reasonably accurate knowledge of parton distribution functions (PDFs), which
describe the structure of the proton in terms of the fraction of its
large longitudinal momentum carried by a quark.
The generalization of this picture from one to three dimensions
(including transverse spatial coordinates)
is a major ongoing effort, to
which significant resources of JLab and CERN are
dedicated, and which is a major science case
for the future electron-ion collider (EIC) \cite{Accardi:2012qut}.

Such 3D hadron structure can be encoded in the generalized parton 
distributions (GPDs) \cite{Mueller:1998fv,Radyushkin:1996nd,Ji:1996nm,Burkardt:2000za},
which are measurable in hard exclusive scattering processes, the most
studied of which is deeply virtual Compton scattering (DVCS)
of a photon with large virtuality $\Q^2$ off a proton,
$\gamma^{*} p \to \gamma p$.
The present phenomenological status (see e.g. \cite{dHose:2016mda,Kumericki:2016ehc})
does not yet allow a reliable determination of most GPDs. We are at the intermediate
stage where one aims for related functions --- Compton form factors (CFFs), which
(at leading order in $1/\Q^2$)
factorize into a convolution of GPDs and the known
perturbatively calculable coefficient functions.
CFFs thus also describe distributions of partons, albeit indirectly, 
while at the same time being more accessible experimentally.
This is completely analogous to the history of DIS studies, where the extraction
of form factors preceded the determination of PDFs.

Since DVCS probes (both the initial virtual and final real photon) couple
to charge, not flavor, 
to determine the distributions of particular quark
flavors it is necessary either to use other processes with flavored
probes (e. g. with meson instead of photon in the final state),
or to combine DVCS measurements with different targets,
like protons and neutrons.
The latter method, involving processes with fewer hadronic states, is less
prone to the influence of low-energy systematic uncertainties, and
will be utilized in this study.

The data on proton DVCS is relatively rich, so
it is the recent complementary \emph{neutron} DVCS measurement by 
JLab's Hall A collaboration \cite{Benali:2020vma}
that made the
present study possible, and enabled us to separate the $u$ and
$d$ quark contributions to the leading CFF.
The Hall A collaboration itself also tried to 
separate $u$ and $d$ quark flavors  in  \cite{Benali:2020vma}, 
using the technique of
fitting separately in each kinematic bin, but their
results are somewhat inconclusive having large uncertainties.
Here, we reduce significantly the uncertainties of
the extracted CFFs by (1) performing global fits and (2) using dispersion
relations (DR), thus imposing additional constraints on CFFs.
To keep our main results model-independent, we parametrize the form factors using
neural networks.

We first perform both model and neural net fits to most of the JLab
6 \GeV{} proton-only DVCS data, 
demonstrating how adding DR constraints
to the neural net procedure significantly increases our ability to extract CFFs. 
We end up with an extraction of six
out of the total eight real and imaginary parts of leading twist-2 CFFs,
including the CFF $\mathcal{E}$, which is a major research target related
to the nucleon spin structure \cite{Ji:1996nm}.
Then, we make both model and neural net fits to JLab's combined proton and 
\emph{neutron} DVCS data. This enables a clear
separation of $u$ and $d$ quark contributions to the leading CFF
$\mathcal{H}$.

\section{Methods and Data}

To connect the sought structure functions to the experimental observables,
we use the formulae from \cite{Belitsky:2001ns,Belitsky:2010jw}, giving 
the four-fold differential cross-section 
$$\frac{d^{4}\sigma_{\lambda,\Lambda}}{d\xb d\Q^2 d|t| d\phi}$$
for the leptoproduction of a real photon by scattering a lepton
of helicity $\lambda/2$ off a nucleon target with longitudinal spin
$\Lambda/2$. 
DVCS is a part of the leptoproduction amplitude and is expressed
in terms of four complex-valued twist-2 CFFs 
$\mathcal{H}(\xi=\xb/(2-\xb), t, \Q^2)$,
$\mathcal{E}(\xi, t, \Q^2)$, $\Ht(\xi, t, \Q^2)$, and $\Et(\xi, t, \Q^2)$.
The kinematical variables are squared momentum transfers from the
lepton, $\Q^2$, and to the nucleon, $t$, Bjorken $\xb$, and the angle $\phi$ between the lepton and photon
scattering planes. The dependence of CFFs on $\Q^2$ is perturbatively
calculable in QCD and will be suppressed in what follows.

An important constraint on CFFs is
provided by dispersion relations \cite{Teryaev:2005uj}, relating
their real and imaginary parts. E. g., for CFF $\mathcal{H}$ we have
\begin{multline}
\ReH(\xi,t) = \Delta(t)  \\
+ \frac{1}{\pi} {\rm P.V.}
\int_{0}^{1} {\rm d}x \left(\frac{1}{\xi-x} - \frac{1}{\xi+x}\right)\ImH (x,t) \;,
\label{eq:DR}
\end{multline}
where $\Delta(t)$ is a subtraction, constant in $\xi$, which is up to an opposite sign
the same for $\mathcal{H}$ and $\mathcal{E}$, and is zero for $\Ht$ and $\Et$. 
This makes it possible
to independently model only the imaginary parts of four CFFs and one subtraction constant.
It is a known feature of any statistical inference that, with given data, 
a more constrained model will generally lead to smaller uncertainties of the results 
--- a property usually called \emph{the bias-variance tradeoff}.
So we expect that, besides easier modeling, the DR constraint will result
in more precise CFFs.
The application of this constraint to neural
network models is the important technical novelty of the fitting procedure presented here.

\subsection{Model fit}
\label{sec:model}
Although the main results of this study are obtained using neural networks, for comparison
we also perform a standard least-squares model fit to the same data.
We use the ``\texttt{KM}''  model parametrization from \cite{Kumericki:2009uq}, which is of a hybrid type: Flavor-symmetric sea quark $H_q$ and gluon $H_G$ GPDs are modeled in the conformal-moment space,
        evolved in $\Q^2$ using leading order (LO) QCD evolution, convoluted with LO coefficient functions,
        and added together to give the total sea CFF $\mathcal{H}^{\rm sea}(\xi, t, \Q^2)$.
On the other hand, valence quark GPDs, like $H^{\rm val}(x,\eta,t)$, are modeled as functions
        of momentum fractions, and only on the $\eta=x$ line, 
        where $x$ and $\eta$  are the average and transferred momentum of the struck parton.
        E.g.,
        \begin{multline}
            H^{\rm val}_{q}(x, x, t) = 
           \frac{n_{q}r_{q}}{1+x}
        \left(\frac{2x}{1+x}\right)^{-\alpha_{v}(t)}
        \left(\frac{1-x}{1+x}\right)^{b_{q}} \\
        \times \frac{1}{
        1-\dfrac{1-x}{1+x}\dfrac{t}{M_{q}^2}} \;, \quad  q = u, d,
            \label{eq:KMval}
        \end{multline}
        where the Regge trajectory $\alpha_{v}(t) = 0.43 + 0.85\,t/\GeV^2$ is used, and
        where the known normalizations $n_q$ of the corresponding PDFs
        are factored out, so that $r_q$ parametrizes ``skewedness''.
        Evolution in $\Q^2$ is neglected for valence CFFs and 
        their imaginary part is given by the LO relation
        \begin{equation}
            \Im\mathcal{H}^{\rm val}(\xi) = \pi\sum_{q=u,d} e_{q}^2 \big[H^{\rm val}_q(\xi,\xi)-
                H^{\rm val}_q(-\xi, \xi) \big] \,,
        \end{equation}
        with $e_{q}$ being the quark charge and where
        dependence on $t$ is suppressed. The subtraction constant
        is modeled separately,
        \begin{equation}
            \Delta(t) = \frac{C}{
            \left(1-\frac{t}{M_{C}^2}\right)^2}\,,
            \label{eq:subtraction}
        \end{equation}
        and real parts of CFFs are then obtained using DR
        (\ref{eq:DR}). $r_q$, $b_q$, $M_q$, $C$ and $M_C$ are parameters of the model.
For further details of the parametrization, and for other CFFs, 
see \cite{Kumericki:2009uq, Kumericki:2015lhb}.
In these references only proton DVCS data, for which the contributions of separate
flavors are not visible, were analysed, so the final model was
designed using simply $H^{\rm val}_u = 2 H^{\rm val}_d$.
Parameters of the model were then fitted to global proton DVCS data resulting in, most recently, the
model \texttt{KM15} \cite{Kumericki:2015lhb}.

In this work we first make a refit of this same model, adding also the 2017 Hall A
data to the dataset, while keeping the same sea parton parameters from the \texttt{KM15} fit
(which was fitted also to H1, ZEUS and HERMES data). The resulting updated fit,
named \texttt{KM20}, is
the only model presented in this paper which is truly global in the sense
that it successfully describes also the low-$x$ H1, ZEUS and HERMES DVCS data.
Then, focusing on flavor separation, we make a fit 
using the same flavor-symmetric sea $\mathcal{H}^{\rm sea}$,
but parametrizing separately $H^{\rm val}_{u}$ and $H^{\rm val}_{d}$ in
(\ref{eq:KMval}), i.e., $r_u \neq r_d$, $b_u \neq b_d$, and $M_u \neq M_d$,
and similarly for other GPDs.
This model is fitted to both proton and neutron DVCS data, where isospin
symmetry is assumed, i. e., we take that 
$H_{d, \mathrm{neutron}} = H_{u, \mathrm{proton}} \equiv H_{u}$, etc.
Since neutron datapoints are few and coming only from JLab, this flavor-separated fit
is performed only to JLab data because only in this kinematics there
is hope to tell flavors apart. The resulting flavor-separated fit is named \texttt{fKM20}.

\subsection{Neural networks fit}

\begin{figure}[t]
    \centering{\includegraphics[width=0.8\linewidth]{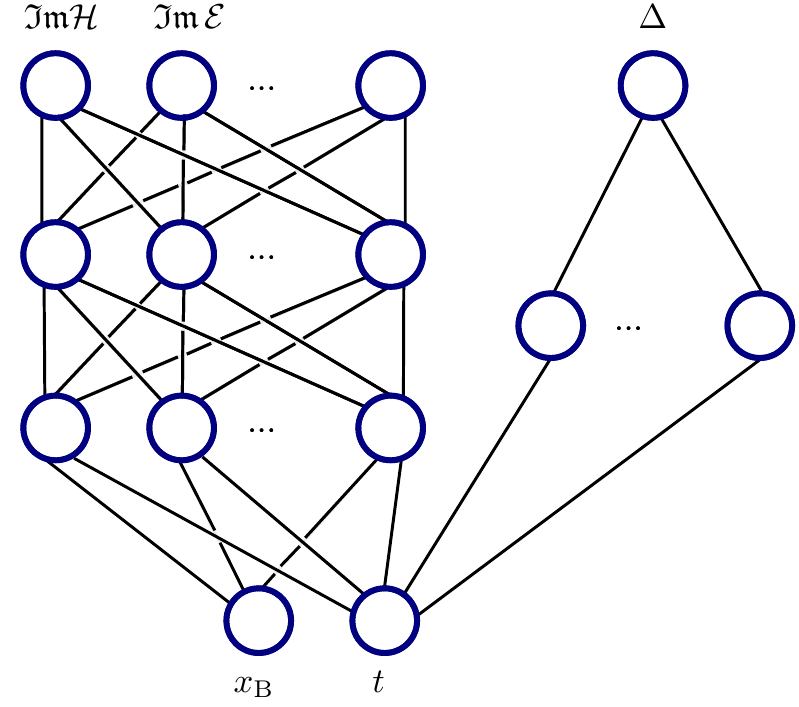}} % Figure image
    \caption{Architecture of neural nets when DR constraints are used. 
        The main net parametrizes imaginary parts of CFFs, 
        while the simpler subsidiary net parametrizes
        the subtraction constant $\Delta(t)$. Real parts are then obtained using DR 
        (\protect\ref{eq:DR}).
        Observables are calculated using total complex CFFs, and, finally, 
        differences with respect to measured observables
    are back-propagated for weight adjustment of the network neurons. There are also standard ``bias'' nodes which are
    for clarity not drawn. Architectures (number of neurons per layer, starting from the input layer) of
    our main nets are $[2\!\to\!13\!\to\!6]$ (for model \texttt{NN20}), $[2\!\to\!13\!\to\!4]$ (\texttt{NNDR20}),
    % $[2\!\to\!11\!\to\!17\!\to\!12]$ (\texttt{fNN20}), 
    and $[2\!\to\!11\!\to\!17\!\to\!8]$ (\texttt{fNNDR20}), 
    while subtraction constant nets are $[1\!\to\!3\!\to\!1]$ (unflavored), $[1\!\to\!5\!\to\!1]$ ($u$-quark), and
    $[1\!\to\!4\!\to\!1]$ ($d$-quark).
} % Figure caption
	\label{fig:architecture} % Label for referencing with \ref{bear}
\end{figure}

For the neural network approach, 
we use the method originally developed by two of us in \cite{Kumericki:2011rz}, and
inspired by a similar procedure for PDF fitting \cite{Forte:2002fg}.
CFFs are parametrized as neural networks, with values at input
representing kinematical variables $\xb$ and $t$, and values at
output representing imaginary or real parts of CFFs.
Here we make significant improvements by adding the possibility of
DR constraints, where outputs represent only imaginary parts,
and one network output represents the subtraction constant $\Delta(t)$ from
(\ref{eq:DR}), see Fig.~\ref{fig:architecture}.
The iterative analysis proceeds in several steps. The network output is                             
used as input for the DR and the result in turn as input for the cross section
formulae. From comparison with experiment we then obtain the required
correction, which is back-propagated to the neural network.
The network parameters are finally adjusted in a standard cross-validated
learning procedure.
For this, we modified the publicly available \textsc{PyBrain} software
library \cite{pybrainpaper}.
This DR-constrained neural net fitting procedure was already
applied by one of us recently to the specific study of 
pressure in the proton \cite{Kumericki:2019ddg},
but is applied here for the first time in a more general context.

To propagate experimental uncertainties, we used the standard
method of fitting to several replicas of datasets \cite{Forte:2002fg}, generated
by Gaussian distributions corresponding to uncertainties of the measured data.
To determine the needed number of replicas to generate and, consequently, the number of neural
nets to train, we made preliminary studies with reduced datasets, where we
compared the results obtained with 10 replicas with those obtained with 80 replicas and we found
that the variation of results is less than 5\%, which we consider acceptable. Since training
of neural nets with DR constraints is quite slow, due to the
evaluation of a numerical Cauchy principal value integral (\ref{eq:DR}) in each training step,
we opted to generate our results with 20 replicas for each presented model.
Training of each net required about 1 day on a single thread CPU of a 2.4 GHz Intel Xeon processor.

Similar preliminary analyses demonstrated that we do not need many neurons to successfully describe
the data, most likely due to the CFF functions being quite well-behaved in this kinematics.
We thus believe that there is no necessity for the \emph{deep learning} with
large amounts of neurons in many layers, which is an extremely powerful method, for
much more complex machine learning tasks. Actual numbers of neurons in our nets
are given in the caption of Fig.~\ref{fig:architecture}.

Preliminary fits using all of the 8 real and imaginary parts of leading
twist-2 CFFs have shown that $\ReHt$ and $\ReEt$ are consistent with zero and
have negligible influence on the goodness of fit, i. e., they cannot
be extracted from the present data.
This is consistent with findings of \cite{Moutarde:2019tqa}, which used an even larger dataset.
Thus, to simplify the model and further reduce the variance, we
removed these two CFFs and performed all neural network fits presented below using just
the remaining six CFFs.

First, to assess the influence of DR, we made fits to the proton-only DVCS data using two parametrizations
\begin{enumerate}
    \item  using neural network parametrization of four imaginary parts of CFFs 
        and of $\ReH$ and $\ReE$ (i.e. without imposing DR constraints) 
     --- this gives us the model \texttt{NN20}.
    \item  using neural network parametrization of four imaginary parts of CFFs, 
        and of the subtraction constant, while
        $\ReH$ and $\ReE$  are then given by DR (\ref{eq:DR}) 
        --- this gives us the model \texttt{NNDR20}.
\end{enumerate}

After we convinced ourselves that the DR-constrained neural net parametrization works in
the proton-only case, we made a
separate parametrization for two light quark flavours (essentially doubling everything)
and fitted to the combined proton and neutron DVCS data --- this gave us the model \texttt{fNNDR20}.

\subsection{Experimental data used}

\begin{table}  % Remove star for one-column table.
\renewcommand{\arraystretch}{1.4}
\caption{\label{tab:chis}
Values of $\chi^2/n_\mathrm{pts}$ for presented models and for
each set of DVCS measurements with fixed proton or neutron
target used in this study ($\phi$-space).
First row specifies the number of real independent CFFs plus the number of
subtraction constants.
Second row gives total value for all datapoints in actually performed fit
(which was just to leading harmonics of Fourier-transformed data --- $n$-space).}
\centering
\setlength{\tabcolsep}{2pt}
\renewcommand{\arraystretch}{1.2}
\begin{tabular}{ccccccc}
\hline\noalign{\smallskip}
 Observable & $n_\mathrm{pts}$ & \texttt{KM20} & \texttt{NN20} & \texttt{NNDR20} 
                            & \texttt{fKM20} & \texttt{fNNDR20} \\
\noalign{\smallskip}\hline
\# CFFs + $\Delta$s &     & 3+1 & 6 & 4+1 & 5+2 &  8+2 \\
\hline
Total (harmonics) & 277 & 1.3 & 1.6 & 1.7 & 1.7 &  1.8 \\
\hline
%%% BEGIN ROWS
CLAS \cite{Pisano:2015iqa}  $A_{\rm LU}$ & 162 & 0.9 & 1.0 & 1.1 & 1.2 & 1.3 \\
CLAS \cite{Pisano:2015iqa}  $A_{\rm UL}$  & 160 & 1.5 & 1.7 & 1.8 & 1.8 & 2.0 \\
CLAS \cite{Pisano:2015iqa}  $A_{\rm LL}$ & 166 & 1.3 & 3.9 & 0.8 & 1.1 & 1.6 \\
CLAS \cite{Jo:2015ema}  $d\sigma$  & 1014 & 1.1 & 1.0 & 1.2 & 1.2 & 1.1 \\
CLAS \cite{Jo:2015ema}  $\Delta\sigma$ & 1012 & 0.9 & 0.9 & 1.0 & 0.9 & 1.1 \\
\hline
Hall A \cite{Defurne:2015kxq}  $d\sigma$  & 240 & 1.2 & 1.9 & 1.7 & 0.9 & 1.3 \\
Hall A \cite{Defurne:2015kxq}  $\Delta\sigma$ & 358 & 0.7 & 0.8 & 0.8 & 0.7 & 0.7 \\
Hall A \cite{Defurne:2017paw}  $d\sigma$  & 450 & 1.5 & 1.6 & 1.7 & 1.9 & 2.0 \\
Hall A \cite{Defurne:2017paw} $\Delta\sigma$ & 360 & 1.6 & 2.2 & 2.2 & 1.9 & 1.7 \\
Hall A \cite{Benali:2020vma}  $d\sigma_{n}$ & 96 &   &   &   & 1.2 & 0.9 \\
\hline
Total ($\phi$-space) & 4018 & 1.1 & 1.3 & 1.3 & 1.2 & 1.3 \\
%%% END ROWS
\noalign{\smallskip}\hline
%\textbf{}
\end{tabular}
\renewcommand{\arraystretch}{1.}
\end{table}

For the neural network fits we used the JLab DVCS data
listed in Table \ref{tab:chis}.
We excluded the lower-$x$ HERA data because we wanted to be safe from any $Q^2$
evolution effects since QCD evolution is not yet implemented in our neural
network framework. Also, in order to demonstrate flavor separation,
it made sense to restrict ourselves to the particular kinematic region where
the neutron DVCS measurement was performed such that there is some balance between the proton and
neutron data.

The data contains measurements of the unpolarized cross-section $d\sigma$, various
beam and target asymmetries defined via
\begin{equation}
    d\sigma_{\lambda,\Lambda}=d\sigma(1+\lambda A_{LU}
    +\Lambda A_{UL} + \lambda\Lambda A_{LL}) \;,
    \label{eq:defA}
\end{equation}
as well as the helicity-dependent cross-section $\Delta\sigma \equiv d\sigma A_{LU}$.
Furthermore, since leading-twist formulae \cite{Belitsky:2001ns,Belitsky:2010jw}
describe observables as
truncated Fourier series in $\phi$, with only one or two terms, we made a Fourier transform
of the data, and fitted only to these first harmonics. 
This makes the fitting procedure much more efficient. 
We propagated the experimental uncertainties using the Monte-Carlo method, and
checked that, indeed, no harmonics beyond the second
one are visible in the data with any statistical significance.

\section{Results and conclusion}

\begin{figure}[t]
    \centering{\includegraphics[width=\linewidth]{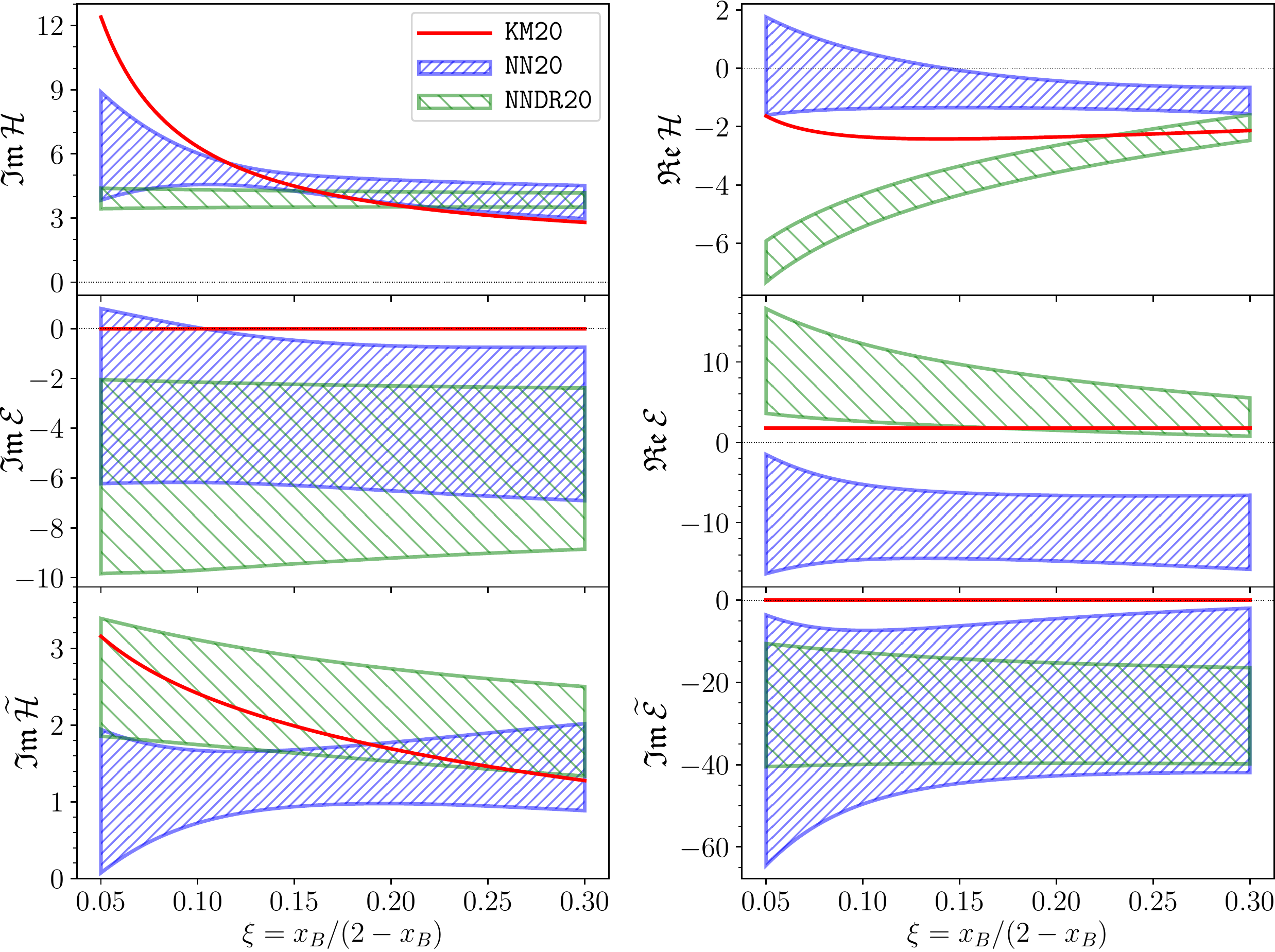}}
    \caption{Extraction of CFFs (at $\Q^2=4\,\GeV^2$ and $t=-0.2\,\GeV^2$) by three fits to JLab proton DVCS data.
    \texttt{KM20} is model described in Sect.~\ref{sec:model},
    \texttt{NN20} is standard neural network parametrization, while \texttt{NNDR20} additionally
includes DR constraints. $\ImE$ and $\ImEt$ are zero in \texttt{KM20} model
by construction.} % Figure caption
	\label{fig:nfDR} % Label for referencing with \ref{bear}
\end{figure}

The quality of the fit for each model is displayed in Table~\ref{tab:chis}.
Judging this quality by the $\chi^2$ values for fits of the above-mentioned 
Fourier harmonics of the data (second row of Table~\ref{tab:chis}) is problematic,
because propagation of experimental uncertainties 
to subleading harmonics is impaired by unknown correlations, see discussion in
Sect. 3.1 of \cite{Kumericki:2016ehc}. We consider the values of $\chi^2$
for the published experimental $\phi$-dependent data as a better measure of the actual fit quality. 
These are displayed in other rows of Table~\ref{tab:chis}.
Some particular datasets are imperfectly described, 
but total values of
$\chi^2/n_{\rm pts}$ of 1.1--1.3 look reasonable and give us confidence that the
resulting CFFs are realistic.
Note that the significantly different number of independent CFFs in our models (see 
the first row of Table~\ref{tab:chis}) leads to a similar quality of fits.
One concludes that there are some correlations among CFFs. Some are intrinsic,
like the consequence of DR, while some will be broken with more data, on more
observables.

\begin{figure}[t]
    \centering{\includegraphics[width=\linewidth]{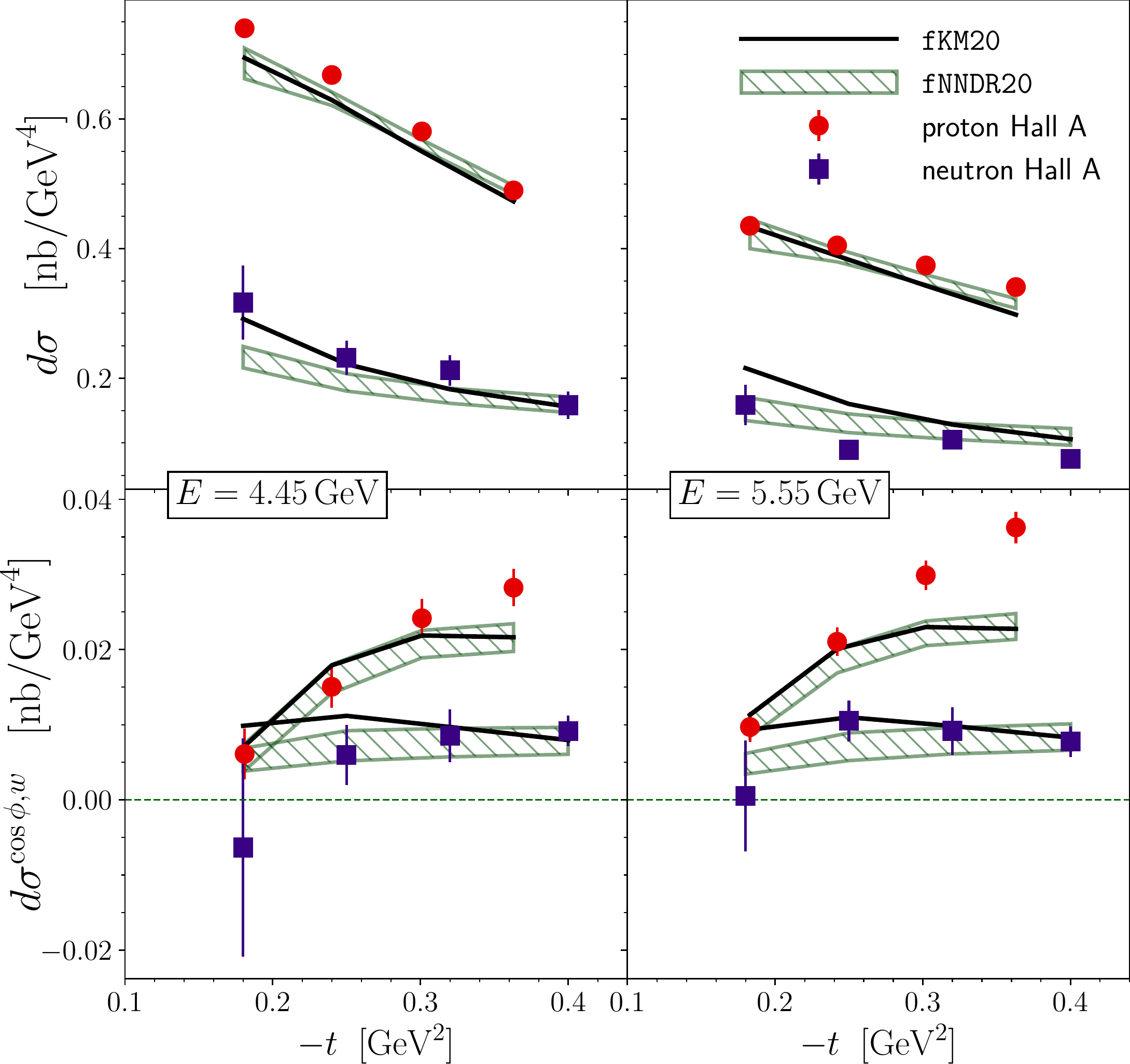}}
    \caption{Model fit \texttt{fKM20} (black solid line) and neural network
        fit \texttt{fNNDR20} (hatched green band) in comparison to Hall A DVCS data on 
        proton (red circles) and neutron (blue squares) cross-sections (upper two
        panels) and first cosine Fourier harmonics of cross-sections (lower two panels), 
        for $x_B = 0.36$, $Q^2 = 1.75\,\GeV^2$, and two beam energies,
        $E=4.45\,\GeV$ (left) and $E=5.55\,\GeV$ (right). 
        % Prediction of cross-sections by the old \texttt{KM15} model is also given (black dotted line).
    } % Figure caption
	\label{fig:pnHallA} % Label for referencing with \ref{bear}
\end{figure}

On Fig.~\ref{fig:nfDR} we display CFFs for models \texttt{KM20},
\texttt{NN20} and \texttt{NNDR20} obtained from fits to proton-only data. We observe
the power of DR constraints, which lead to reduced uncertainties of the \texttt{NNDR20}
model in comparison to \texttt{NN20},
most notably for $\ImH$, $\ReH$, and $\ImEt$.
Mean values are also shifted for real parts of unpolarized CFFs $\mathcal{H}$ and
$\mathcal{E}$, where DR constraints can even change the sign of the extracted CFFs\footnote{Interestingly,
    the popular VGG \cite{Vanderhaeghen:1999xj} and GK \cite{Goloskokov:2007nt} 
    models have negative $\ReE$,
while the fit in \cite{Moutarde:2018kwr} gives a positive $\ReE$ in this region.}. For $\ReH$,
DR induce a strong $\xi$ dependence and a clear extraction of this CFF. It is this
particular effect of DR that made recent attempts at determination of quark
pressure distribution in the proton from the DVCS data possible \cite{Burkert:2018bqq,Kumericki:2019ddg}.
The DR-constrained neural net fit \texttt{NNDR20} is, as is to be expected, 
in somewhat better agreement with the model fit \texttt{KM20} which is also DR constrained.
The green bands in the second row of Fig.~\ref{fig:nfDR} constitute the first unbiased extraction 
of the important CFF $\mathcal{E}$ in this kinematic region.

The CFFs in the \texttt{NNDR20} model are in broad agreement with
the results of the (also DR-constrained) 
model fit of Ref. \cite{Moutarde:2018kwr}. One notable exception is the opposite
sign of $\ImE$. As this CFF has the largest uncertainty of the six displayed, 
one can hope that with more data the discrepancy will fade.
Comparing with CFFs extracted by the recent global neural network fit of Ref. \cite{Moutarde:2019tqa},
results agree within specified uncertainties, with the largest tension now being observed for $\ReE$.
One notes that $\ReE$ of \cite{Moutarde:2019tqa} agrees much better with our
fit \texttt{NN20}, which is to be expected since one is now comparing results of more
similar procedures: both are completely unbiased fits, with neither using DR
constraints.

\begin{figure}[t]
    \centering{\includegraphics[width=\linewidth]{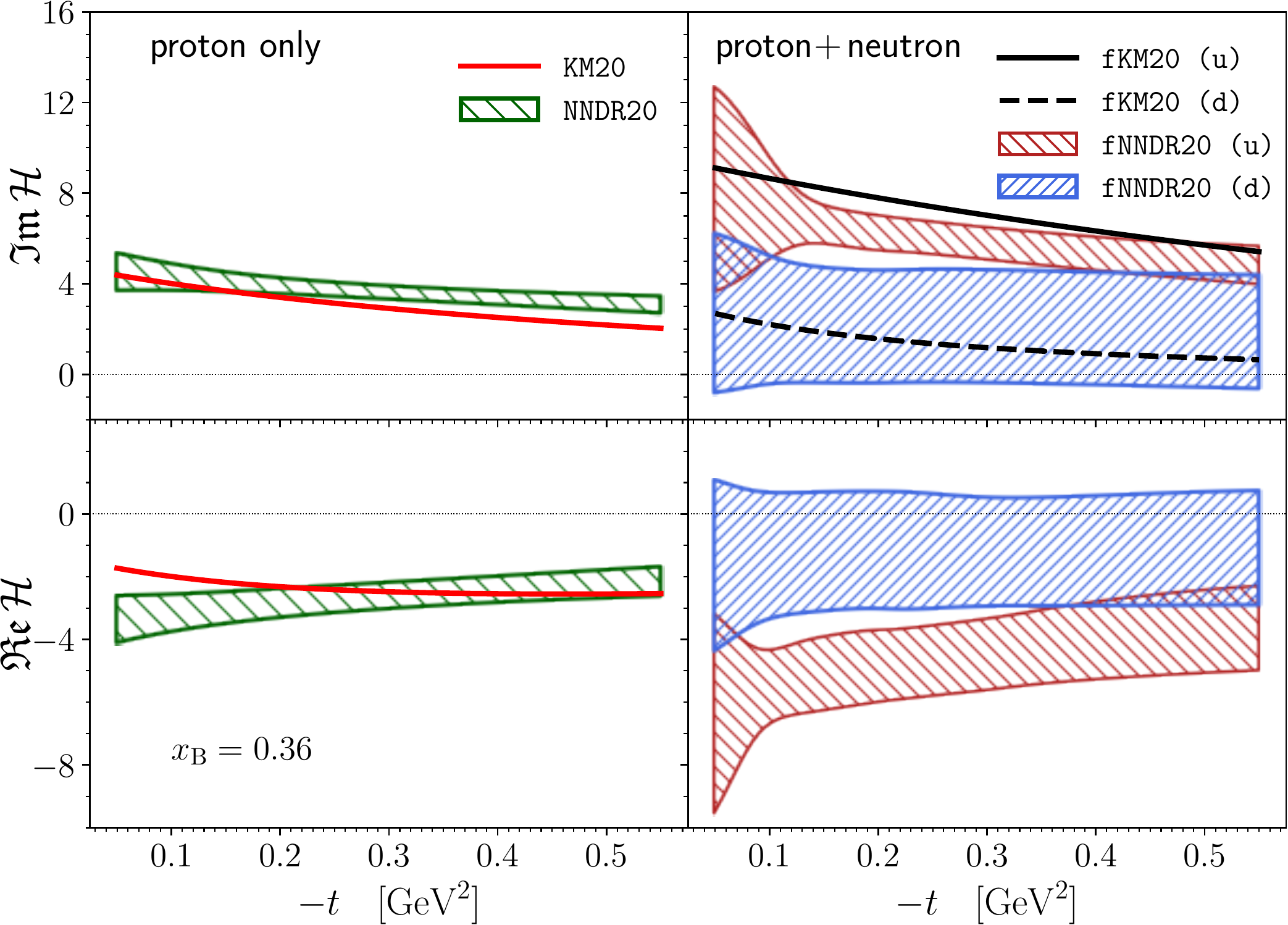}}
    \caption{CFF $\mathcal{H}$ extracted (at $\Q^2=4\, \GeV^2$ and $x_{\rm B} = 0.36$) from neural network fit to proton-only DVCS data (left). Separation of $u$ (red band) and $d$ (blue band) quark CFF $\mathcal{H}$ resulting from neural network fit to proton and neutron JLab DVCS data (right). Red solid
        (unflavored), black solid ($u$) and
        dashed ($d$) lines correspond to analogous least-squares model fit to the same data.
    } % Figure caption
	\label{fig:fCFFH} % Label for referencing with \ref{bear}
\end{figure}

\begin{figure}[t]
    \centering{\includegraphics[width=\linewidth]{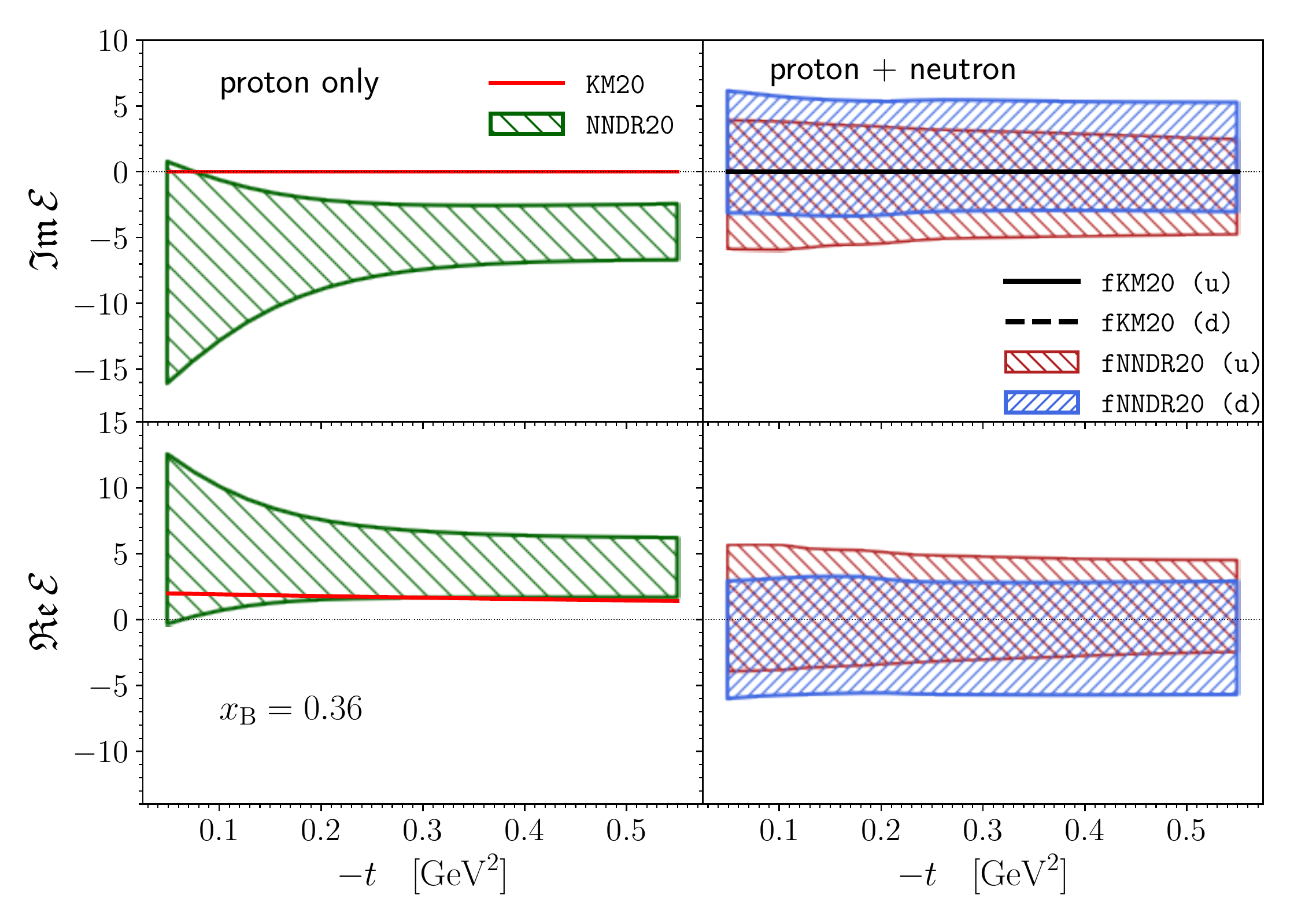}}
    \caption{Same as Fig.~\protect\ref{fig:fCFFH} but for CFF $\mathcal{E}$. Separation of $u$ and $d$ 
        quark CFF $\mathcal{E}$ is not possible with present data (right). $\ImE$ is zero
        in \texttt{KM} models by construction.
} % Figure caption
	\label{fig:fCFFE} % Label for referencing with \ref{bear}
\end{figure}

Turning now to the simultaneous fit to proton and neutron data, besides in Table~\ref{tab:chis},
the quality of the fit can also be seen in Fig. \ref{fig:pnHallA},
where the model fit \texttt{fKM20} and the DR-constrained neural net fit \texttt{fNNDR20}
are confronted with Hall A data.
The resulting $\ImH$ and $\ReH$ CFFs, separately for up and down quarks,
are displayed in the right two panels of Fig.~\ref{fig:fCFFH}, demonstrating
how the inclusion of neutron DVCS data enables a clear flavor separation
for this CFF. (For other CFFs, there is no visible separation, see 
example of $\mathcal{E}$ in Fig.~\ref{fig:fCFFE}.)

The separated up and down quark CFFs have much larger uncertainties than
their sum, shown in the left panels of Fig.~\ref{fig:fCFFH}, and although
there are some hints of different $t$-slopes, at this
level we are not yet able to address the question of possible different
spatial distributions of up and down quarks in the nucleon.

To conclude, we have used JLab DVCS data to make both a model-dependent and an unbiased
neural net extraction of six Compton form factors, where constraints by dispersion relations
proved valuable. Furthermore, in the case of the dominant CFF $\mathcal{H}$,
we have successfully separated the contributions of up and down quarks. 
This constitutes another
step towards a full three-dimensional picture of the nucleon structure.

\subsection*{Code availability}
In the interest of open and reproducible research, the computer code used in
the production of numerical results and plots for this paper is made available at 
\url{https://github.com/openhep/neutron20}.

%----------------------------------------------------------------------------------------
%	ACKNOWLEDGEMENT
%----------------------------------------------------------------------------------------

\subsection*{Acknowledgments}
This publication is supported by the
Croatian Science Foundation project IP-2019-04-9709, by
Deutsche Forschungsgemeinschaft (DFG) through the
Research Unit FOR 2926, ``Next Generation pQCD for Hadron Structure: Preparing
for the EIC'', project number 40824754,
by QuantiXLie Centre of Excellence through grant KK.01.1.1.01.0004, and the 
EU Horizon 2020 research and innovation programme, STRONG-2020 project, 
under grant agreement No 824093.

%----------------------------------------------------------------------------------------
%	BIBLIOGRAPHY
%----------------------------------------------------------------------------------------

%\bibliography{kkumer.bib}
%\bibliographystyle{apsrev}

%----------------------------------------------------------------------------------------

\end{document}